\documentclass{aastex}
\usepackage{psfig}
\usepackage{emulateapj5}
\usepackage{apjfonts}
\newcommand{\msun}{\mbox{M$_{\odot}$}}

\newcommand{\Pbif}{$P_{\mathrm{bif}}$}

\newcommand{\swin}{Centre for Astrophysics and Supercomputing, Swinburne
University of Technology, P.O.~Box 218 Hawthorn, VIC 3122, Australia}

\newcommand{\etal}{et al.\ }

\begin{document}
\title{Recycled Pulsars Discovered at High Radio Frequency}
\author{R. T. Edwards and M. Bailes}
\affil{\swin}
\begin{abstract}
We present the timing parameters of nine pulsars discovered in a
survey of intermediate Galactic latitudes at 1400 MHz with the Parkes
radio telescope.  Eight of these pulsars possess small pulse periods
and period derivatives thought to be indicative of ``recycling''.  Six
of the pulsars are in circular binary systems, including two with
relatively massive white dwarf companions. We discuss the implications
of these new systems for theories of binary formation and evolution.
One long-period pulsar (J1410--7404) has a moderately weak magnetic
field and an exceedingly narrow average pulse profile, similar to other
recycled pulsars.
\end{abstract}


\section{Introduction}
Since the early days of pulsar astronomy it has been known that 
%we are dealing with 
radio pulsars are steep-spectrum objects \citep{ccl+69}. The majority of pulsar
surveys were conducted at low frequency to exploit the higher flux and
increased telescope beam size made available.  By the mid 1990s
70-cm surveys had resulted in the discovery of $\sim$700 pulsars,
approximately 60 of which belonged to the class of ``millisecond
pulsars'' (MSPs; $P\lesssim 25$ ms).  However the elevated sky
temperature and effects of interstellar dispersion,
scatter-broadening and terrestrial interference present significant
barriers to the discovery of pulsars, particularly MSPs, at low
frequency. Early efforts at 20-cm surveys \citep{stwd85,cl86,jlm+92}
were surprisingly successful in discovering large numbers of pulsars,
however the time resolution afforded by the backend systems did not
provide sufficient sensitivity to discover any MSPs. We have conducted
a high frequency survey of intermediate Galactic latitudes with the
new multibeam 21-cm receiver and pulsar backend at the Parkes
radio telescope. The rapid sky coverage afforded by the multiple beams
combined with improved backend hardware has enabled us to overcome the
limitations of previous surveys and succeed in discovering a large
number of millisecond pulsars.  In this paper we discuss nine new %
low-magnetic-field pulsars found in this survey.

\section{The Swinburne Intermediate Latitude Pulsar Survey}
\label{sec:survey}
 Over a period of 12 months from 1998 August - 1999 August, we
conducted a pulsar survey with the 13-beam 21-cm ``multibeam''
receiver at the Parkes radio telescope \citep{swb+96}. The rapid sky
coverage and high sensitivity of this system makes for a formidable
survey instrument. Coupled with the thirteen $2\times96$-channel
filterbanks covering 288 MHz each centred at a frequency of 1374~MHz,
pulsars are being discovered at a rapid rate.  The pulsar backend was
originally built for an ongoing deep survey of the Galactic plane
which is expected to almost double the known pulsar population
\citep{lcm+00,clm+00}. Based on Monte Carlo simulations similar to
those discussed by \citet{tbms98}, we found that 5--10 previously
unknown millisecond pulsars ought to be detectable with unprecedented
time efficiency in a shallower survey flanking the region of the
Galactic plane survey. With the surfeit of computational power made
available with the upgrade of the Swinburne supercomputer to 64 Compaq
Alpha workstations, we chose to use half the sampling interval of the
Galactic plane survey (that is, 125~$\mu$s vs. 250~$\mu$s) in the hope
of discovering any nearby sub-millisecond pulsars.  With 265-s
integrations covering the region enclosed by $5\degr < |b| < 15\degr$
and $-100\degr < l < 50\degr$ we completed the survey in 14 days of
integration time and discovered $\sim70$ pulsars\footnote{Some
candidates are yet to be confirmed} pulsars of which 8 show spin-down
behavior consistent with recycling, confirming the predictions of the
simulations. The survey was sensitive to slow and most millisecond
pulsars with flux densities greater than approximately $0.3$--$1$
mJy.  Full details of the survey will be available in a forthcoming
paper (R.T.~Edwards \etal 2001, in preparation).

\section{New Recycled Pulsars}
\label{sec:newmsps}

\begin{figure*}
\centerline{\psfig{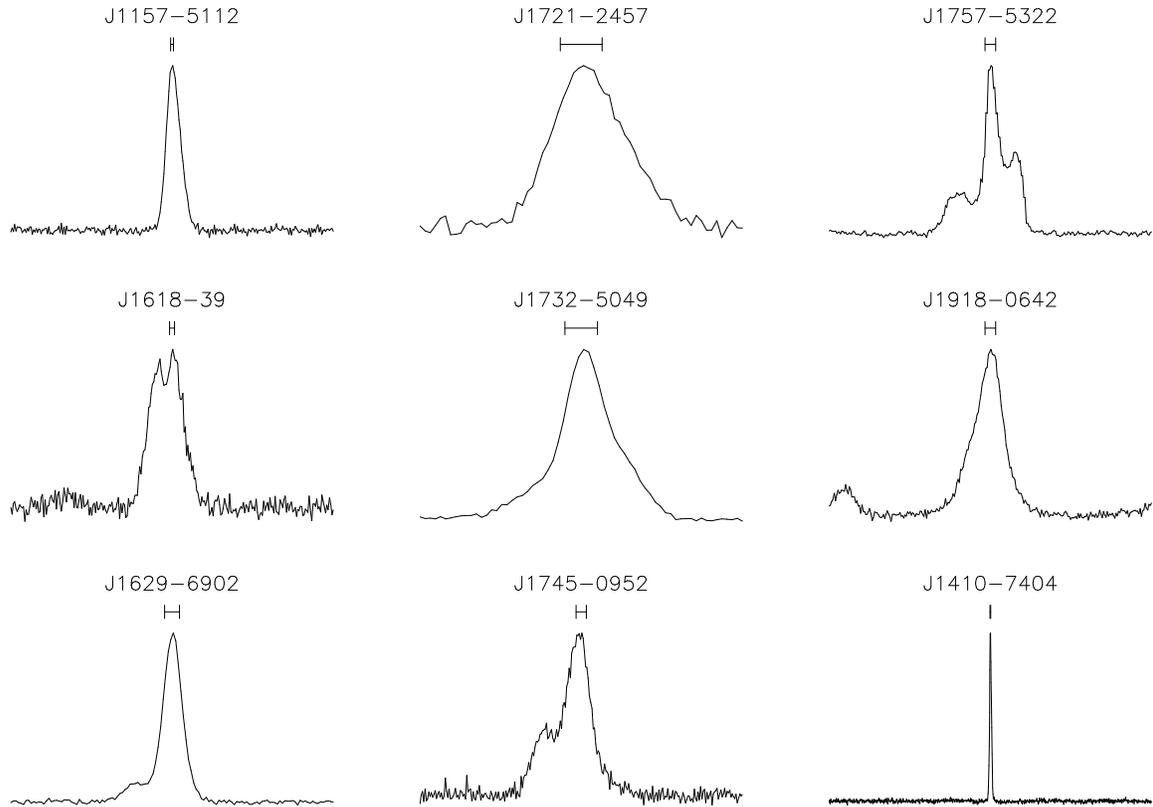}}
\caption{Average pulse profiles for the pulsars presented in this
paper.  These result from the summation of several observations using
the system described in \S \ref{sec:survey} (except for the
J1618--39 profile which arises from data taken with a $2\times512\times0.5$
MHz channel system at a center frequency of 1390 MHz).  The vertical
flux scale is arbitrary and normalized between pulsars.  Horizontal
bars represent the degree of time smearing induced by the differential
dispersion delay within individual filterbank channels.}
\label{fig:profs}
\end{figure*}

Pulsars discovered in the intermediate latitude survey were subjected
to a campaign of pulsar timing observations to determine their
astrometric, spin and orbital parameters. The center beam of the
multibeam system was used to make observations of typically 250
seconds in length. The raw filterbank samples were folded offline at
the topocentric pulse period to form integrated pulse profiles that
were subsequently fitted to a ``standard'' profile to produce a
nominal pulse time-of-arrival (TOA).  These TOAs were then used to fit
for the parameters of a standard timing model using the TEMPO software
package\footnote{http://pulsar.princeton.edu/tempo} with the DE200
solar system ephemeris \citep{sta82}. To avoid the high degree of
covariance between the time of periastron and the longitude of
periastron experienced with nearly circular binaries, we used the ELL1
binary timing model \citep{lcw+01} which fits for the time of
ascending node and the Laplace parameters $e \cos \omega$ and $e \sin
\omega$ instead. In the absence of such covariance the uncertainty
estimates of the ELL1 model are more reliable, however its result is
subject to the assumption of a very small eccentricity.  To check the
validity of this assumption, particularly in the case of PSR
J1157--5112 which has a large eccentricity compared to most white
dwarf pulsar binaries, we have also fitted for the BT model
\citep{bt76} and find the parameters consistent to better than $0.1
\sigma$. Due to its long orbital period, we have not yet been able to
obtain a phase-connected solution for PSR J1618--39\footnote{The name
of this pulsar may change when a better position measurement becomes
available.}; the parameters presented derive from analysis of secular
variations in barycentric pulse period.

The parameters of nine newly discovered pulsars with small pulse
periods and/or spin-down rates are listed in Tables 1--3. Values in
parentheses represent the error in the last quoted digit, calculated
from twice the formal uncertainty produced by TEMPO.  Dispersion
measures were fit for with the inclusion of one or more TOAs produced
from 660~MHz data, which in the absence of adequate signal to noise
ratio were obtained by cross-correlation with the same template
profile used for the high frequency TOAs. For this reason the errors
quoted are probably underestimated. Also presented in the table are
several derived quantities of interest, including the minimum
companion mass for binary systems assuming a 1.35 \msun\ companion
\citep{tc99}, the surface magnetic field strength assuming a dipole
geometry (given by $B_{\rm surf} \simeq 3.2\times 10^{19}$ Gauss
s$^{-1/2} \sqrt{P \dot{P}}$), the ``characteristic age'' assuming
magnetic dipole spin-down (with a braking index of 3) from a very fast
initial rotation and the distance and Galactic $z$-height derived from
the electron density model of \citet{tc93} (accurate on average to
around 30\%).  Average pulse profiles from these pulsars are provided
in Figure \ref{fig:profs}. As is visible in the plot, PSR J1918--0642
possesses an interpulse which follows the main pulse by $\sim190\degr$
with $\sim15$\% of its peak intensity. Apparent in the profile of
PSR J1618--39 is a weak component leading the main pulse by $\sim120\degr$
with $\sim10$\% of the peak intensity.

These systems are of particular interest because they most likely
belong to the class of so-called ``recycled'' pulsars. In the standard
model, the origin of the short spin period and low inferred magnetic
field strength of these pulsars lies in an earlier epoch of binary
interaction \cite[e.g.][]{bv91}.  Matter accreted from the companion
simultaneously spins up the pulsar and dramatically attenuates its
magnetic field. The resultant rapid rotation rate provides the
required conditions for pair production (even with a ``weak'' magnetic
field) and the pulsar, which had previously ceased to emit after
spinning down below the critical rate, is recycled into a radio-loud
short-period pulsar.

\begin{figure*}
\centerline{\psfig{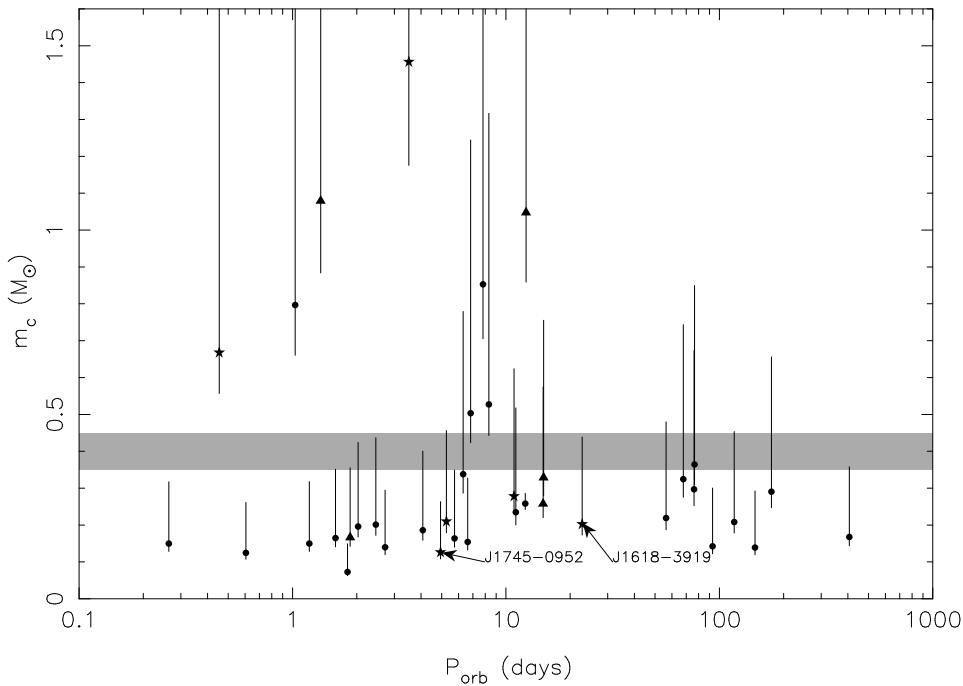}}
\caption{Distribution in orbital period and companion mass of known
circular-orbit pulsars in the Galactic disk (excluding those with
planets or evaporating companions). Points and upper and lower ends of
error bars represent median, 90\% upper limit and minimum companion
masses assuming randomly inclined orbits and a 1.35 \msun\ pulsar,
with the exception of PSR B1855+09 which has much smaller errors
resulting from a Shapiro delay measurement \citep{ktr94}. Stars,
triangles and circles represent pulsars discovered in the intermediate
latitude survey (this work), the Galactic plane survey \citep{clm+01},
and previously known systems respectively.  The gray bar between
0.35--0.45 \msun\ represents the transition from He to CO white dwarf
companions.}
\label{fig:pbm2}
\end{figure*}

Of the pulsars listed in Tables 1--3, three are isolated and six are
members of binary systems. Of the six binaries, four are members of a
large class of millisecond pulsars with low mass white dwarf
companions (\S \ref{sec:lmbp}). The remaining two are orbiting massive
white dwarfs (\S \ref{sec:imbp}) in orbits sufficiently close that
relativistic perturbations to the timing behavior should be measurable
in a few years. PSR J1757--5322 in particular emits sufficient
gravitational radiation that its orbit will decay to the point of
coalescence within a Hubble time, with dramatic and unknown
consequences \citep{eb01a}. The systems were discovered with signal to
noise ratios in the range of 12--45, implying flux densities of
approximately 0.4--8 mJy and luminosities of 0.8--60 mJy kpc$^2$,
comparable to those of previously known MSPs in the Galactic
disk, although a thorough treatment must await calibrated flux
density measurements.

There does not seem to be any particular preference for high or low
Galactic latitudes in the pulsars presented here. A Kolmogorov-Smirnov
test on the seven new ``millisecond'' pulsars ($P <$ 20~ms) indicates
that the observed values of $|b|$ differ from a uniform distribution
at a significance level of only 18\%. Clearly a campaign to survey
higher Galactic latitudes with the multibeam system would also be
worthwhile in terms of millisecond pulsar discoveries. The dearth of
MSPs found so far in the Galactic plane survey \citep{mlc+00} is
puzzling in light of the lack of any indication of diminishing returns
from our survey at smaller latitudes with the same observing
system. Although we employ a higher sampling rate, the dispersion
smearing induced by the filterbank channels is greater than two sample
intervals for all new MSPs. In addition, it was discovered (after
completion of the survey!) that the strength of a ubiquitous and
problematic signal generated by the sampler is much reduced at
slower sample rates.

\section{The Formation of Low Mass Binary Pulsars}
\label{sec:lmbp}

The newly discovered pulsar PSR J1618--39 lies in the middle of what
previously appeared to be a ``gap'' \citep{cam95} in the distribution
of binary pulsar orbital periods (see Figure \ref{fig:pbm2}). Whilst
its spin-down rate has not been measured, only a recycled pulsar or a
very young pulsar could exhibit such rapid rotation, and its large
displacement from the Galactic plane ($\sim$650 pc) rules out the
latter option. Since it is recycled and its orbital parameters other
than orbital period are typical of the low mass binary pulsars
(LMBPs), we contend that the physics of the formation of LMBPs do not
exclude such orbital periods. With the discovery of this pulsar and
two (J1732--5049, J1918--0642), perhaps three (J1745--0952 appears to
have an anomalously long rotation period; see \S \ref{sec:imbplong}) other
LMBPs, it is timely to reassess the standard model of their
formation.

Briefly, the standard evolutionary scenario for LMBPs (see
e.g. \citealt{pk94}) is as follows: a pair of main-sequence stars
($\sim$10 \msun\ and $\sim$1 \msun\ in mass respectively) are in a
binary system. The (more massive) primary evolves to fill its Roche
lobe and begins unstable mass transfer in a common envelope (CE)
phase. The envelope is eventually expelled, the remaining core of the
primary collapses in a supernova and a pulsar is formed. The pulsar
shines for several Myr, after which the pulsar has spun-down to the
extent that radio emission ceases. The secondary eventually evolves to
fill its Roche lobe, transferring matter (with angular momentum) to
the neutron star, ``recycling'' it by spinning it up to millisecond
periods where emission is once again possible (despite the
accompanying large reduction in magnetic field strength). It is
believed that the low mass X-ray binaries are in fact systems in this
phase. For systems with orbital periods greater than 1--2 d at the
beginning of this phase the orbit is expected to expand to a final
period $\ga$ 50 d due to the transfer of mass from a lighter secondary
to a heavier primary. However, for closer initial orbits, angular
momentum losses due to magnetic braking and gravitational radiation
become important and the orbit will shrink during the mass transfer
stage. Hence a ``gap'' in the orbital period distribution is expected
from around this critical ``bifurcation'' period, \Pbif\ $\simeq 2$~d
\citep[e.g.][]{ps88} to $\sim$50~d. 

\begin{figure*}
\centerline{\psfig{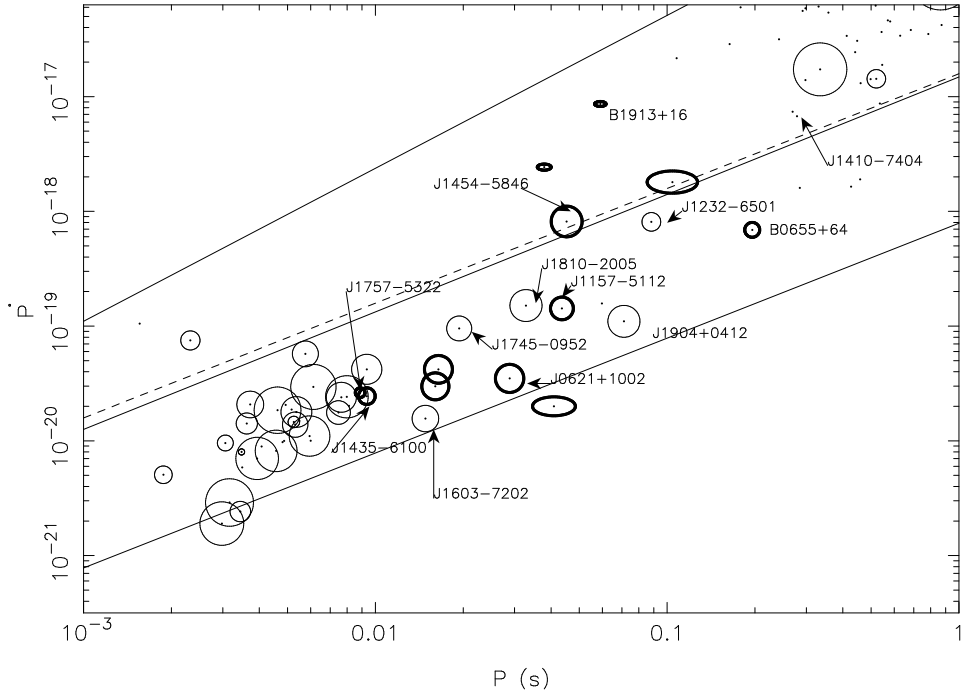}}
\caption{Distribution of known disk pulsars in the recycled region of the
$P$-$\dot{P}$ plane (excluding systems with planets or evaporating
companions). Solid lines represent (from top to bottom) the expected
positions of pulsars at $t=0$, 1 and 20 Gyr after spin-up to 
equilibrium with an Alfv\'en-terminated accretion disk. The dashed line
depicts a line of constant ``characteristic age'' (i.e. assuming infinite
initial spin frequency) at $\tau_{\rm c} = 1$ Gyr.
After \citet{acw99}, points representing binary pulsars
are enclosed by a circle with a radius proportional to 
1+$\log P_{\mathrm orb}$ (d). Lines are bold in the case of
massive companions, and elliptical in the case of eccentric binaries.
}
\label{fig:ppdot}
\end{figure*}

Figure \ref{fig:pbm2} shows the distribution of orbital periods and
companion masses for all known circular orbit pulsar binaries in the
Galactic disk. There is arguably still an under-density of pulsars
between 12~d $< P_{\rm orb} <$ 56~d, even including the newly discovered
PSR J1618--39 and the 22.7-d low-mass binary PSR J1709+23 \citep{cad97}
which is not depicted in Figure \ref{fig:pbm2} due to the lack of a
published mass function.  We note that the discovery of just one
system with $P_{\rm orb} \simeq 45$ d would be sufficient to remove
the suggestion of an under-density, especially if the massive systems
around $P_{\rm orb}=10$ d are disregarded due to their probable
different origin.  Even if the under-density is real, the presence of
significant numbers of systems between the bifurcation period and
$P_{\rm orb}=12$~d requires explanation.  \citet{tau96} argues that
systems with initial orbital periods only slightly longer than \Pbif\ may
experience moderate angular momentum losses that are sufficient to
reduce the degree of spiral-out. To preserve the observed underdensity
in this case one must require that the significance of the magnetized
stellar wind is a sharp function of orbital separation, with the
transition from moderate to insignificant magnetic braking occuring
at some orbital period greater than \Pbif.

Two further models predict the presence of pulsars in the range 1~d $<
P_{\rm orb} <$ 12~d. The first begins with a binary system consisting
of a pulsar and a star on the red (or asymptotic) giant branch, and
produces a recycled pulsar in a close binary via a common envelope
spiral-in phase. The remnant companion is either a helium dwarf or, in
the case of Roche lobe overflow on the asymptotic giant branch, a CO
dwarf with $m_{\rm c} >$ 0.45 \msun. Doubt has been cast over the
plausibility of this model for many binary pulsars due to energy
considerations \citep{tkr00}, however for several pulsars such as the
newly discovered PSR J1157--5112 it is the only feasible scenario (see
\S \ref{sec:imbp}).  An alternative model that may produce systems in
the required orbital period range is that of ``massive case A'' or ``early
case B'' mass transfer \citep{tkr00,tvs00}.  In this scenario a massive
star on the late main sequence or early red giant branch loses much of
its envelope in a phase of highly super-Eddington mass transfer, and a
common envelope phase is avoided due to the radiative envelope of the
companion, which responds to mass loss by contracting. The end result
is a CO white dwarf in excess of 0.35~\msun\footnote{In the work of
\citet{tvs00} the scenario occasionally produces a ligher He dwarf.}
in an orbit shorter than $\sim$70~d.  This is the model favored by
\citet{tkr00}, however it is clear from Figure \ref{fig:pbm2} that
most of the systems with 1~d $<P_{\rm orb} <$ 12~d must be lighter He
dwarfs. We expect that to account for this high proportion of He
dwarfs the standard LMBP model must be capable of producing some
systems in this orbital period range, perhaps via the curtailment of
spiral-out by moderate angular momentum losses \citep{tau96}.

\section{Intermediate Mass Binary Pulsars}
\label{sec:imbp}
With the discovery of PSRs J1157--5112, J1757--5322 (\citealt{eb01a};
this paper), J1435--6100 and J1454--5846 \citep{clm+01}, there are now
eight recycled pulsars known to have heavy (CO or ONeMg) white dwarf
companions (see Table
\ref{tab:imbp}).  As mentioned in the previous section, there are two
plausible scenarios for the formation of heavy white dwarfs in close
orbits: massive late case A/early case B mass transfer
\citep{tvs00,tkr00} or common envelope evolution on the first
\citep{tau96} or second \citep{vdh94} ascent of the red giant
branch. The first scenario is limited to systems with orbital periods
greater than a few days (thus excluding J1757--5322, B0655+64, J1435--6100
and perhaps J1232--6501)
and companions lighter than $\sim$0.9~\msun (thus excluding
J1157--5112 and perhaps J1454--5846). 
The common envelope scenario is able to produce the
systems excluded from the first scenario, albeit with some difficulty
for J1157--5112.
In the common envelope formalism of
\citet{web84}, one requires $\eta\lambda > 2$ to reproduce the present
orbital separation of the J1157--5112 system (where $\lambda \approx
0.5$ depends on the stellar density distribution and affects the
estimate of its binding energy and $\eta$ is the
efficiency of use of orbital energy in envelope expulsion). In the
absence of other plausible models, we must conclude for now that the
system did evolve through a CE phase and that either additional energy
sources are available ($\eta > 1$) or the binding energy of the
stellar envelope is lower than is generally assumed. The latter option
has some support from recent numerical studies \citep{dt00}.

Figure \ref{fig:ppdot} shows a region of the $P$--$\dot{P}$ diagram
for all presently known disk pulsars. With the recent discoveries made
at Parkes the region between 10 and 100~ms is now well populated. We
note that the positive correlation visible in the plot is simply a
result of magnetic spin down over time scales necessarily shorter than
the age of the Galaxy. It is clear that the standard assumption of
relatively short initial spin period and magnetic dipole spindown with
constant magnetic field strength is in error for the millisecond
pulsars due to the anomalously large ages inferred, especially when
kinematic contributions to the observed $\dot{P}$ are taken into
account \citep[e.g.][]{ctk94}. Indeed, even when the initial spin
frequency is limited to that allowed in the standard spin-up scenario,
little impact is made on the inferred ages. In Figure \ref{fig:ppdot}
we adopt a spin-up line defined by $\dot{P_0} = 1.1 \times 10^{-15}
{\rm s}^{-4/3} P_0^{4/3}$ \citep{acw99} and show the predicted
positions of pulsars immediately after spin-up, at 1 Gyr and at 20
Gyr. Also shown is a dashed line representing a ``characteristic age''
of 1 Gyr assuming $P_0 \ll P$, tracing virtually the same locus in the
$P$--$\dot{P}$ diagram.

%\begin{figure}
\vspace{8mm}
\centerline{\psfig{figure=pbe.ps,width=3.5in}}
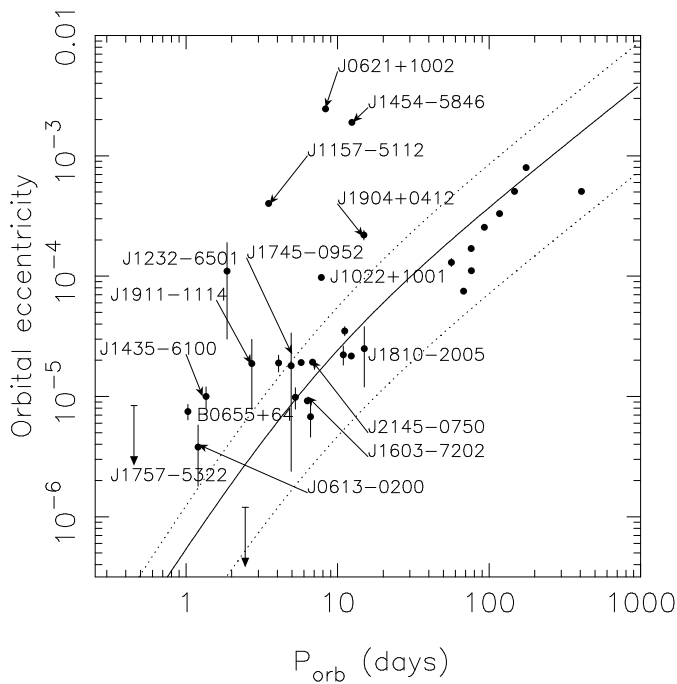
\figcaption{Orbital periods and eccentricities of circular pulsar binaries
in the Galatic disk. The solid and upper and lower dashed curves 
represent the median and 95\% upper and lower bounds respectively of 
an orbital period -- eccentricity relation based on that of
\citet{phi92b}, modified for applicability to all orbital periods
(S. Phinney 2000 pers. comm.). Error bars in general represent twice the
nominal uncertainty produced by TEMPO. It should be noted that the
ordinate has a logarithmic scale and that larger error bounds would make
some measurements consistent with zero.
\label{fig:ecc}}
%\end{figure}
\vspace{8mm}

In the work of
\citet{acw99}, massive pulsar binaries were divided into two groups,
the ``B1913+16-like'' systems and the ``J0621+1002-like''
systems. These were distinguished by a tendency in the 0621 group for
wider orbits and more complete recycling (i.e. smaller periods and
particularly period derivatives).  In the context of this scheme the
newly discovered systems J1435--6100 and (especially) J1757--5322 are
anomalous by virtue of their strongly recycled nature and close
orbits.

\subsection{Systems with Intermediate Spin Periods}
\label{sec:imbplong}

In the past it seemed that nearly all LMBPs had $P < 10$ ms whilst all
IMBPs had longer spin periods.  It has been argued mainly on this basis
\citep{ts99a,acw99} that PSR J1603--7202 ($P=15$ ms) also possesses a
CO companion in a relatively face-on orbit, hence its low mass
function.  The newly discovered moderately recycled pulsars
J1810--2005, J1904+0412 and especially J1745--0952 and J1232--6501 also
have small mass functions despite their relatively long spin periods,
suggesting perhaps a similar interpretation. However with the
discovery of IMBPs J1435--6100 and J1757--5322, both with
$P\sim$~9~ms, it is clear that there is some overlap between LMBPs and
IMBPs in $P$-$\dot{P}$ space and that assignments to either class
based on these parameters are subject to significant uncertainty. In
addition, it is quite unlikely {\it a priori} for this many systems to
be as face on as is required to make them IMBPs: only around one in 20
pulsar binaries with $0.45$~\msun\ companions would be expected to
have mass functions as low as those of J1745--0952 or J1232--6501.

If the newly discovered recycled systems with $P >10$~ms and small
mass functions are ordinary LMBPs that have evolved from LMXBs, it is
difficult to see why only one of the previously known LMBPs rotated so
slowly. The relative insensitivity of the multibeam backend to pulsars
with $P \lesssim 10$~ms at dispersion measures typical of recycled
pulsars discovered in the deep Galactic plane survey ($\sim
200$~cm$^{-3}$~kpc; \citealt{clm+01}) can account for some but
probably not all of the high proportion of recent recycled discoveries
with $P >10$~ms. Indeed, it could be argued that the pulse periods of
the four new low mass systems reported here are also anomalous,
and for this sample dispersion smearing imposes no more of a selection
effect than in the previous surveys that detected the known population.
A Kolmogorov-Smirnov test finds their distribution different to
that of previously known LMBPs with 91\% significance, however
with a sample size of four this conclusion should be taken with caution.

One possibility is that the long period systems
evolved instead from massive late case A / early case B mass transfer
\citep{tvs00} resulting in a low mass He companion.  \citet{clm+01}
favor this interpretation and suggest that such systems were
preferentially detected in the Galactic plane and intermediate
latitude surveys due to the low scale height of this population in the
Galactic disk. Whilst no conclusions should be drawn without full
population modelling, in the work of \citet{tvs00} it seems that most
systems produced by this channel ought to possess heavy CO
companions. Since there are five known systems with $P >10$~ms and
small mass functions that one may suggest evolved through this
channel, one would expect to have discovered many more than the four
remaining known CO systems that are allowed by this scenario.

One may look to orbital eccentricities to resolve questions of
evolutionary history, since in the absence of a final stage of stable
mass transfer IMBPs are not expected to obey the LMBP orbital period
-- eccentricity relation of \citet{phi92b} \citep{cnst96}. The
situation has become more complicated (Figure \ref{fig:ecc}) with
updated measurements of previously known pulsars \citep{tsb+99} and
newly discovered pulsars (this work; \citealt{clm+01}).  Since PSR
J2145--0750 has a minimum companion mass consistent with a CO
companion in an orbit as circular as most LMBPs, we cannot take the
similar eccentricities of PSRs J1603--7202, J1745--0952 and
J1810--2005 as evidence against a massive companion.  Likewise the
probable high eccentricities (relative to the relation) of J0613--0200
and J1911--1114, pulsars which we have no reason to doubt are LMBPs,
mean that we must be conservative in our interpretation of candidate
IMBPs on the basis of large eccentricities, particularly at short
orbital periods --- a connection one might be tempted to make in the
case of J1904+0412 and especially J1232--6501 due to their positions
in Figures \ref{fig:ppdot} and \ref{fig:ecc}.

Optical identifications of large numbers of recycled pulsar companions
would provide useful information about the statistical correlation
between observable pulsar parameters and companion mass by means of
temperature measurements in the context of white dwarf cooling models.
The newly discovered systems add significantly to the sample potentially
detectable with 8-m class telescopes.

\section{PSR J1410--7404: Disrupted Binary?}

As shown in Figure \ref{fig:profs}, the mean pulse profile of the
newly discovered isolated pulsar J1410--7404 is extraordinarily
narrow.  Given its period of 279~ms, the pulse width -- period
relation of \citet{ran90} predicts a minimum width of 4.6$^\circ$
(FWHM), double the measured value of 2.3$^\circ$. Rankin's model fits
normal pulsars well and the only exceptions to it (more than a
factor of few percent smaller than the predicted minimum) in the 688
published pulse widths recorded in our local pulsar catalogue are the
recycled pulsars for which there appears to be no correlation between
period and duty cycle (see also \citealt{kxl+98}). The period derivative of
J1410--7404 is $6.8 \times 10^{-18}$, again a very small value
relative to other pulsars with slow spin periods.  This leads us to
suggest that the pulsar may in fact be recycled.  From the period and
period derivative we infer a magnetic field strength of $4.4 \times
10^{10}$~G, lower than all but five apparently un-recycled pulsars and
comparable to the double neutron star systems B1913+16 (at $2.3 \times
10^{10}$~G) and B1534+12 (at $1.0 \times 10^{10}$~G). A wider orbit
and/or unfavorably oriented kick could easily result in the disruption
of the orbit of a system similar to the progenitors of B1913+16 or
B1534+12 at the time of the companion's supernova \citep{bai89},
leading to an isolated mildly recycled pulsar similar to
J1410--7404. Unlike the double neutron star systems that coalesce due
to gravitational radiation while still possessing pulsars with
relatively fast spin periods, such systems would continue to be
observable as ``slow'' pulsars with weak magnetic fields throughout
their spin-down to the pair production death line.

Of the five slow pulsars now known with weaker inferred magnetic
fields, PSRs J1320--3512 \citep{dsb+98}, B1952+29 and B1848+04
\citep{gl98} do not possess narrow profiles. Should either of the
remaining two slow pulsars (J1355--6206 and J1650--4341, reported
at the web site of the Galactic plane survey 
collaboration\footnote{http://www.atnf.csiro.au/$\sim$pulsar/psr/pmsurv/pmwww/}) 
with weak magnetic fields possess narrow average
profiles we suggest a similar possible interpretation to J1410--7404.
It is possible that other weakly magnetized pulsars with long spin
periods are also recycled but possess broader pulse profiles, due to
the dependence of pulse width on $\alpha$, the angle between the
pulsar spin and magnetic axes. This dependence could be removed
through polarimetric measurements of $\alpha$, as in \citet{kxl+98}.

\acknowledgements We would like to thank T.M. Tauris for useful
discussions and the Parkes Multibeam pulsar survey team for making the
parameters of new pulsars available on the internet prior to formal
publication. We are grateful for the assistence given by W. van
Straten and M. Britton in making observations, and to the Parkes
Multibeam team for exchanging telescope time to provide more frequent
timing observations. The authors appreciate the insightful and helpful
comments of the referee, F. Camilo.

%\bibliographystyle{apj1}
%\bibliography{journals_apj,local,modrefs,psrrefs,grbrefs,crossrefs}

\singlespace

\begin{deluxetable}{lrrr}
\tablecaption{Astrometric, Spin, Binary and Derived Parameters}
\tablecolumns{4}
\tablewidth{0pt}
\tablehead{
    \colhead{} 
      & \colhead{J1157--5112}
      & \colhead{J1410--7404}
      & \colhead{J1618--39}
}
\startdata
Right ascension, $\alpha$ (J2000.0)\dotfill 
   & 11$^{\mathrm h}$57$^{\mathrm m}$08\fs166(1)
   & 14$^{\mathrm h}$10$^{\mathrm m}$07\fs370(5)
   & 16$^{\mathrm h}$18$^{\mathrm m}$30(50)$^{\mathrm s}$
\\
Declination, $\delta$ (J2000.0)\dotfill
& --51\degr12\arcmin56\farcs14(3)
& --74\degr04\arcmin53\farcs32(2)
& --39\degr19(10)\arcmin%       10\farcs
\\
Pulse period, $P$ (ms)\dotfill
   & $43.58922706284$(12)
   & $278.7294436271$(15)
   & $11.987313$(5)
\\
$P$ epoch (MJD)\dotfill
    & 51400.0
    & 51460.0
    & \nodata
\\
Period derivative, $\dot{P}$ (10$^{-20}$)\dotfill
   & $14.3$(10)
   & $674$(9)
    & \nodata
\\
Dispersion Measure (pc cm$^{-3}$)\dotfill
   & $39.67$(4)
   & $54.24$(6)
   & $117.5$(4)
\\
Orbital Period, $P_{\rm orb}$ (d)\dotfill
   & $3.50738640$(3)
    & \nodata
   & $22.8$(2)
\\
Projected semi-major axis, $a \sin i$ (lt-s)\dotfill
   & $14.28634$(3)
    & \nodata
   & $10.24$(17)
\\
Epoch of Ascending Node, $T_{\rm asc}$ (MJD)\dotfill
   & $51216.4442640$(13)
    & \nodata
   & $51577.37$(8)
\\
$e \cos \omega$\dotfill
   & $-0.000323$(4)
    & \nodata
    & \nodata
\\
$e \sin \omega$\dotfill
   & $0.000240$(4)
    & \nodata
    & \nodata
\\
Span of timing data (MJD) \dotfill &
51197--51888 & 51309--51889 & \nodata
\\
Weighted RMS timing residual ($\mu$s)\dotfill&
73 & 60 & \nodata
\\
Pulse width at FWHM, $w_{\rm 50}$ (\degr)\dotfill&
18 & 2.3 & 43
\\
Pulse width at 10\% peak, $w_{\rm 10}$ (\degr)\dotfill&
33 & 4.5 & \nodata
\\
\cutinhead{Derived Parameters}
Longitude of periastron, $\omega$ (\degr)\dotfill
   & $306.6$(6)
    & \nodata
    & \nodata
\\
Orbital eccentricity, $e$\dotfill
   & $0.000402$(4)
    & \nodata
    & \nodata
\\
Minimum Companion mass, $m_{\rm c}$ (\msun)\dotfill
    & 1.18
    & \nodata
    & 0.18
\\
Characteristic age, $\tau_c$ (Gyr)\dotfill
    & 4.8
    & 0.65
    & \nodata
\\
Surface magnetic field, $B_{\rm surf}$ (10$^8$ Gauss)\dotfill
    & 25
    & 440
    & \nodata
\\
Galactic longitude, $l$ (\degr)\dotfill
    & 294.3
    & 308.3
    & 340.8
\\
Galactic latitude, $b$ (\degr)\dotfill
    & $10.75$
    & $-12.04$
    & $7.88$
\\
Distance, $d$ (kpc)\dotfill
    & 1.9
    & 2.1
    & 4.8
\\
Distance from Galactic plane, $|z|$ (kpc)\dotfill
    & 0.35
    & 0.45
    & 0.65
\\
\enddata
\end{deluxetable}
\newpage
\begin{deluxetable}{lrrr}
\tablecaption{Astrometric, Spin, Binary and Derived Parameters}
\tablecolumns{4}
\tablewidth{0pt}
\tablehead{
    \colhead{} 
      & \colhead{J1629--6902}
      & \colhead{J1721--2457}
      & \colhead{J1745--0952}
}
\startdata
Right ascension, $\alpha$ (J2000.0)\dotfill 
   & 16$^{\mathrm h}$29$^{\mathrm m}$08\fs7706(4)
   & 17$^{\mathrm h}$21$^{\mathrm m}$05\fs496(2)
   & 17$^{\mathrm h}$45$^{\mathrm m}$09\fs1400(9)
\\
Declination, $\delta$ (J2000.0)\dotfill
& --69\degr02\arcmin45\farcs294(3)
& --24\degr57\arcmin06\farcs1(4)
& --09\degr52\arcmin39\farcs67(5)
\\
Pulse period, $P$ (ms)\dotfill
   & $6.0006034432179$(19)
   & $3.496633727625$(7)
   & $19.37630312709$(3)
\\
$P$ epoch (MJD)\dotfill
    & 51600.0
    & 51600.0
    & 51500.0
\\
Period derivative, $\dot{P}$ (10$^{-20}$)\dotfill
   & $1.00$(3)
   & $0.59$(10)
   & $9.5$(4)
\\
Dispersion Measure (pc cm$^{-3}$)\dotfill
   & $29.490$(3)
   & $47.758$(19)
   & $64.474$(14)
\\
Orbital Period, $P_{\rm orb}$ (d)\dotfill
    & \nodata
    & \nodata
   & $4.9434534$(2)
\\
Projected semi-major axis, $a \sin i$ (lt-s)\dotfill
    & \nodata
    & \nodata
   & $2.378615$(17)
\\
Epoch of Ascending Node, $T_{\rm asc}$ (MJD)\dotfill
    & \nodata
    & \nodata
   & $51270.674499$(8)
\\
$e \cos \omega$\dotfill
    & \nodata
    & \nodata
   & $10.0$(148)$\times 10^{-6}$
\\
$e \sin \omega$\dotfill
    & \nodata
    & \nodata
   & $1.5$(16)$\times 10^{-5}$
\\
Span of timing data (MJD) \dotfill &
51395--51889 & 51395--51889 & 51254--51889
\\
Weighted RMS timing residual ($\mu$s)\dotfill&
7.3 & 56 & 72
\\
Pulse width at FWHM, $w_{\rm 50}$ (\degr)\dotfill&
24 & 95 & 33
\\
Pulse width at 10\% peak, $w_{\rm 10}$ (\degr)\dotfill&
60 & \nodata & 186
\\
\cutinhead{Derived Parameters}
Longitude of periastron, $\omega$ (\degr)\dotfill
    & \nodata
    & \nodata
   & $34$(49)
\\
Orbital eccentricity, $e$\dotfill
    & \nodata
    & \nodata
   & $1.8$(16)$\times 10^{-5}$
\\
Minimum Companion mass, $m_{\rm c}$ (\msun)\dotfill
    & \nodata
    & \nodata
    & 0.11
\\
Characteristic age, $\tau_c$ (Gyr)\dotfill
    & 9.5
    & 9.4
    & 3.2
\\
Surface magnetic field, $B_{\rm surf}$ (10$^8$ Gauss)\dotfill
    & 2.5
    & 1.5
    & 13.8
\\
Galactic longitude, $l$ (\degr)\dotfill
    & 320.4
    & 0.4
    & 16.4
\\
Galactic latitude, $b$ (\degr)\dotfill
    & $-13.93$
    & $6.75$
    & $9.90$
\\
Distance, $d$ (kpc)\dotfill
    & 1.4
    & 1.6
    & 2.4
\\
Distance from Galactic plane, $|z|$ (kpc)\dotfill
    & 0.33
    & 0.18
    & 0.41
\\
\enddata
\end{deluxetable}
\newpage
\begin{deluxetable}{lrrr}
\tablecaption{Astrometric, Spin, Binary and Derived Parameters}
\tablecolumns{4}
\tablewidth{0pt}
\tablehead{
    \colhead{} 
      & \colhead{J1732--5049}
      & \colhead{J1757--5322}
      & \colhead{J1918--0642}
}
\startdata
Right ascension, $\alpha$ (J2000.0)\dotfill 
   & 17$^{\mathrm h}$32$^{\mathrm m}$47\fs7671(4)
   & 17$^{\mathrm h}$57$^{\mathrm m}$15\fs1615(3)
   & 19$^{\mathrm h}$18$^{\mathrm m}$48\fs0363(8)
\\
Declination, $\delta$ (J2000.0)\dotfill
& --50\degr49\arcmin00\farcs11(1)
& --53\degr22\arcmin26\farcs387(5)
& --06\degr42\arcmin34\farcs80(5)
\\
Pulse period, $P$ (ms)\dotfill
   & $5.312550204595$(6)
   & $8.869961227277$(4)
   & $7.645872761379$(18)
\\
$P$ epoch (MJD)\dotfill
    & 51575.0
    & 51570.0
    & 51600.0
\\
Period derivative, $\dot{P}$ (10$^{-20}$)\dotfill
   & $1.38$(9)
   & $2.63$(4)
   & $2.4$(3)
\\
Dispersion Measure (pc cm$^{-3}$)\dotfill
   & $56.839$(9)
   & $30.817$(7)
   & $26.595$(17)
\\
Orbital Period, $P_{\rm orb}$ (d)\dotfill
   & $5.26299721$(6)
   & $0.4533112381$(5)
   & $10.9131774$(5)
\\
Projected semi-major axis, $a \sin i$ (lt-s)\dotfill
   & $3.982868$(5)
   & $2.086527$(5)
   & $8.350489$(16)
\\
Epoch of Ascending Node, $T_{\rm asc}$ (MJD)\dotfill
   & $51396.3659935$(12)
   & $51394.1080693$(3)
   & $51569.117366$(4)
\\
$e \cos \omega$\dotfill
   & $2.9$(19)$\times 10^{-6}$
   & $-1.3$(42)$\times 10^{-6}$
   & $-1.3$(4)$\times 10^{-5}$
\\
$e \sin \omega$\dotfill
   & $-9.4$(20)$\times 10^{-6}$
   & $3.8$(44)$\times 10^{-6}$
   & $-1.8$(4)$\times 10^{-5}$
\\
Span of timing data (MJD) \dotfill &
51306--51889 & 51306--51889 & 51395--51889
\\
Weighted RMS timing residual ($\mu$s)\dotfill&
17 & 21 & 40
\\
Pulse width at FWHM, $w_{\rm 50}$ (\degr)\dotfill&
58 & 17 & 35
\\
Pulse width at 10\% peak, $w_{\rm 10}$ (\degr)\dotfill&
136 & 81 & \nodata
\\
\cutinhead{Derived Parameters}
Longitude of periastron, $\omega$ (\degr)\dotfill
   & $163$(12)
   & $341$(61)
   & $216$(11)
\\
Orbital eccentricity, $e$\dotfill
   & $9.8$(20)$\times 10^{-6}$
   & $4.0$(44)$\times 10^{-6}$
   & $2.2$(4)$\times 10^{-5}$
\\
Minimum Companion mass, $m_{\rm c}$ (\msun)\dotfill
    & 0.18
    & 0.56
    & 0.24
\\
Characteristic age, $\tau_c$ (Gyr)\dotfill
    & 6.1
    & 5.3
    & 5.0
\\
Surface magnetic field, $B_{\rm surf}$ (10$^8$ Gauss)\dotfill
    & 2.8
    & 4.9
    & 4.3
\\
Galactic longitude, $l$ (\degr)\dotfill
    & 340.0
    & 339.6
    & 30.0
\\
Galactic latitude, $b$ (\degr)\dotfill
    & $-9.45$
    & $-13.98$
    & $-9.12$
\\
Distance, $d$ (kpc)\dotfill
    & 1.8
    & 1.4
    & 1.4
\\
Distance from Galactic plane, $|z|$ (kpc)\dotfill
    & 0.30
    & 0.33
    & 0.22
\\
\enddata
\end{deluxetable}

\newpage

\begin{deluxetable}{lccccl}
\tablewidth{0pt}
\tablecolumns{6}
\tablecaption{Possible Intermediate Mass Binary Pulsars}
\tablehead{
 \colhead{} & \colhead{$P$} & \colhead{$\dot{P}$} &
    \colhead{$P_{\mathrm orb}$} & \colhead{$m_{\rm c}$\tablenotemark{1}} &
    \colhead{}\\
 \colhead{Name}&\colhead{(ms)} & \colhead{} & \colhead{(d)} &\colhead{(\msun)}
 & \colhead{Ref}
}
\startdata
% Following produced by ppdot and -'s changed to --'s manually
% and added refs and sorted
J0621+1002 & 28.85 & 3.5 $\times 10^{-20}$ & 8.32 & 0.53 & 4\\ 
B0655+64 & 195.7 & 6.9 $\times 10^{-19}$ & 1.03 & 0.78 & 1\\ 
J1022+1001 & 16.45 & 4.2 $\times 10^{-20}$ & 7.81 & 0.85 & 4\\ 
J1157--5112 & 43.59 & 1.5 $\times 10^{-19}$ & 3.51 & 1.47 & 6\\ 
J1232--6501 & 88.28 & 7.9 $\times 10^{-19}$ & 1.86 & 0.17 & 5\\ 
J1435--6100 & 9.348 & 2.3 $\times 10^{-20}$ & 1.35 & 1.08 & 5\\ 
J1454--5846 & 45.25 & 8.0 $\times 10^{-19}$ & 12.42 & 1.05 & 5\\ 
J1603--7202 & 14.84 & 1.6 $\times 10^{-20}$ & 6.31 & 0.34 & 3\\ 
J1745--0952 & 19.38 & 9.2 $\times 10^{-20}$ & 4.94 & 0.126 & 7\\ 
J1757--5322 &  8.87 & 2.8 $\times 10^{-20}$ & 0.45 & 0.67 & 6\\ 
J1810--2005 & 32.82 & 1.3 $\times 10^{-19}$ & 15.01 & 0.33 & 5\\ 
J1904+0412 & 71.09 & 1.0 $\times 10^{-19}$ & 14.93 & 0.26 & 5\\ 
J2145--0750 & 16.05 & 3.0 $\times 10^{-20}$ & 6.84 & 0.50 & 2\\ 

\enddata
\tablenotetext{1}{Assuming an orbit inclined at 60\degr\ and a 1.35 \msun\ pulsar}
\tablerefs{
(1) \citealt{dth78};
(2) \citealt{bhl+94};
(3) \citealt{llb+96}
(4) \citealt{cnst96};
(5) \citealt{clm+01};
(6) \citealt{eb01a};
(7) this work
}
\label{tab:imbp}
\end{deluxetable}

\end{document}